*Michael Leznik[1]*

# Hubs and Authorities of the English Premier League for 2010-2011

## Keywords

In this work author applies well known web search algorithm Hyperlink – Induced Topic Search (HITS) to problem of ranking football teams in English Premier League (EPL). The algorithm allows the ranking of the teams using the notions of hubs and authorities well known for ranking pages in the World Wide Web. Results of the games introduced as a graph where losing team 'gives a link' to a winning team and, if draw registered both team give links to each other. In case of a win link is weighted as three points in adjacent matrix and in case of draw as one point. Author uses notion of authority in order to define team which win a game and hub as a team which lose a game, the winner of the competition defined as the 'worst' hub, team that didn't reinforced any other team. Using this ranking system, the champion's team, which is a 'worst hub' must not lose, or draw games to other 'good authorities' teams. If by the end of the competition there are teams with an equal number of wins and losses then the team which has beaten more teams with higher authority ranks, wins.

---

[1] The author works as Data Scientist in King.com, London office (some of this research has been written while author worked as Chief research Scientist in Greenlight, London)

**Preliminaries**

An algorithm to recognise hubs and authorities (HITS)[2] in the World Wide Web (WWW) environment has been developed by [1]. The target pursued was to develop an algorithm, which allows extracting information from the link structures of web environments. In its turn such information would provide a tool for finding relevant to topic web pages and rank them. The algorithm ranks pages by assigning two numbers to each *relevant* page; a hubs and authority weights. Numbers are discovered recursively, which means in a process of repeating itself, like for instance two parallel mirrors staying in front of each other will recursively reflect each other an infinite number of times. However, unlike the mirrors' reflections, programming recursions are stopped after a sufficient number of iterations and not run infinitely. At the end of the calculations one would obtain two sets of numbers where hub weights would identify pages pointing to many pages with high authority and authority weights in their turn will identify the pages pointed to by pages with high hub weights. One can see that in the same way that two parallel mirrors reflect each other, pages with high authority enforce pages with high hub value and vice versa. HITS is widely, but probably mistakenly, considered to be a precursor for PageRank, developed by [2], most probably both algorithms were developed in parallel as it happens quite often in science and, if this article was about the history of science I could have provided a curious reader with at least a dozen of examples of such parallel discoveries.

In order to find a set of pages with higher hub and authority values [1] suggests first selecting an initial set of pages relevant to a query, such set of initial pages we shall call *a query space* . Selection can be performed based on the content of the pages under consideration, where content relevancy can be identified by use of one of the Natural Language Processing algorithms, for instance Latent Semantic Analysis (LSA), for explanation of LSA see [3]. In his experiments in order to find pages semantically relevant to a query string Kleinberg used the top 200 pages returned by search engines such as Alta Vista or Hotbot. However, as mentioned above the same result might be obtained by running your own analysis, based on LSA, or other appropriate Natural Language Processing algorithm. Initially the selected set will not contain many reciprocally connected pages, nevertheless as [1] points out one has to start somewhere. In order to find more related (connected) pages an initial seed is augmented by pages which point to the set of initial pages and any other pages which is pointed to by the initial set of pages. This way we should increase the number of strong authorities in the initial seed by expanding it along the links that enter and leave it.

---

[2] HITS stands for Hyperlink – Induced Topic Search

After obtaining and augmenting the initial seed of pages we can start to calculate hub and authority values. However, before we move on to numbers let us review it graphically, after all, as [1] points out the algorithm operates on focused subgraphs of the WWW that we construct from the text-based output of a search engine; our technique for constructing such subgraphs is designed to produce small collections of pages likely to contain the most authoritative pages for a given topic.

Anyone who is familiar with link structures of web pages will probably confirm an assumption that authoritative pages will have a lot of incoming links from common pages and hub pages would have a lot of outgoing links to common pages. Thus, for instance pages containing the results of the recent football matches in English Premier League (EPL) would have a lot of incoming links from pages discussing these results, while in their turn EPL results pages would have a lot of outgoing links to those participating in the league competition e.g. clubs, referring authorities, merchandising companies and so on. It is also safe to assume that different authorities of the same query space will rarely link to each other mainly because of the competition between them. Thus, for instance, one football club will not link to another football club in order to acknowledge its authority however hubs will freely link to different authorities due to their cross topic features; The BBC news might link to different football clubs pages from the page containing article about football game between those clubs. Based on the said we safely conclude that hubs are pointing out to authorities while authorities, in referral, point out to hubs, see Figure 1 .

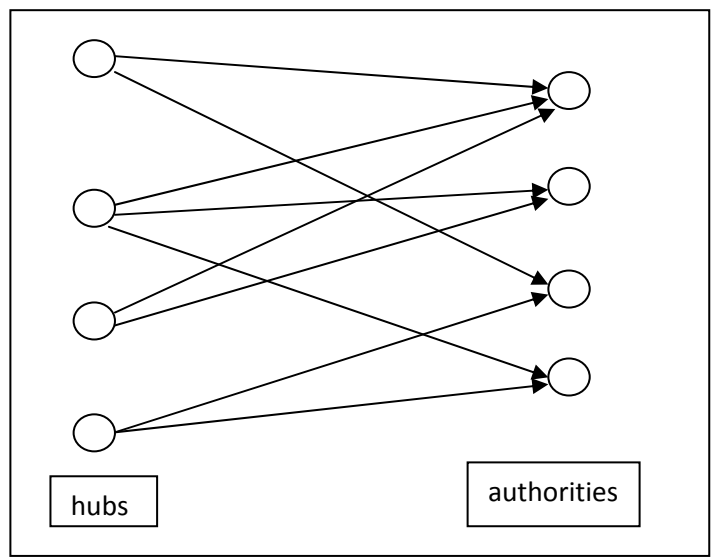

Figure 1

Now let us turn to numbers and explain all what was said above from the mathematical point of view. With each page $p$ we associate a non negative authority weight $x^{(p)}$ and non negative hub weight $y^{\langle p \rangle}$. In order to keep on equal scaling all weights are normalized such that the sum of squares of the corresponding weights equals to one.

$$\sum \left(x^{\langle p \rangle}\right)^2 = 1 \text{ and } \sum \left(y^{\langle p \rangle}\right)^2 = 1$$

Numerically, it is natural to express the mutually reinforcing relationship between hubs and authorities as follows:

$$x^{\langle p \rangle} \leftarrow \sum y^{\langle p \rangle} \text{ authority weights and}$$

$$y^{\langle p \rangle} \leftarrow \sum x^{\langle p \rangle} \text{ hub weights}$$

In plain English it means that, if $p$ points to many pages with large $x$ values then it should receive a large $y$ value; and if $p$ pointed to by many pages with large $y$ values, then it should receive a large $x$ value, for graphical depiction see Figure 2 and Figure 3.

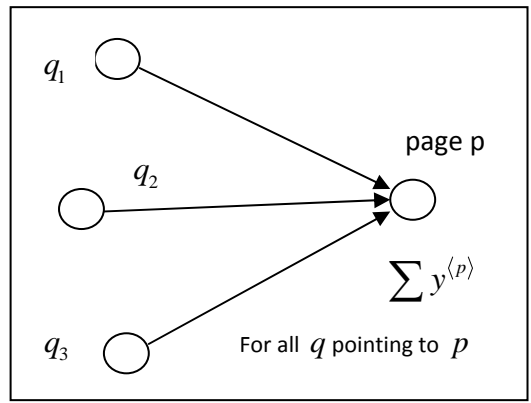

**Figure 2**

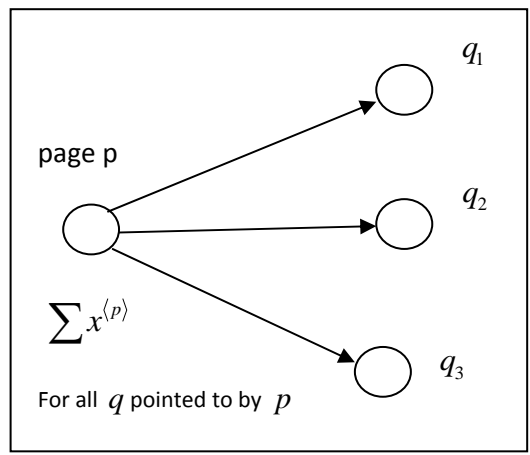

**Figure 3**

The mathematical procedure behind these calculations is called Eigenanalysis; where the German word eigen means peculiar in English. Detailed information on Eigenanalysis can be found in [4], [5][3]. In order to demonstrate the process in detail let us discuss the following toy example. Assume we have a four nodes web graph with the following relations, see Figure 4.

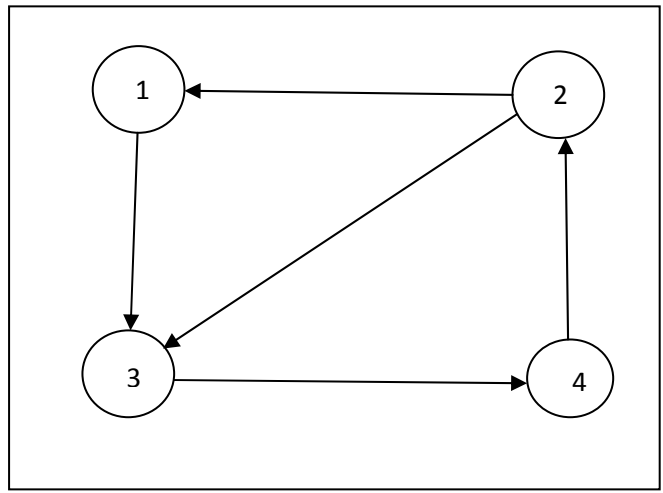

Figure 4

First we have to build an adjacency matrix which reflects the relationships between nodes.

$$\mathbf{A} = \begin{bmatrix} 0 & 0 & 1 & 0 \\ 1 & 0 & 1 & 0 \\ 0 & 0 & 0 & 1 \\ 0 & 1 & 0 & 0 \end{bmatrix}$$

Rows and columns of the matrix are nodes in our webgraph and numbers show their relationships. Thus for instance the number 1 in the first row, third column means that there is a link from node 1 to node 3. Numbers in the second row, in columns one and three, identify links from node 2 to nodes 1 and 3. In the same manner the third row will identify links from node 3 to node 4 and the fourth row link from node 4 to node 2. Performing Eigenanalysis on the matrix $\mathbf{A'A}$ we obtain the following results for the authority weights:

$$\mathbf{a} = \begin{bmatrix} 0.52 \\ 0 \\ 0.85 \\ 0 \end{bmatrix}$$

And after performing the same operation on the matrix $\mathbf{AA'}$ we obtain results for the hub weights:

---

[3] The only two books I mention here are probably my favourite, but one can find another at least dozen, if not more books and articles on the topic.

$$\mathbf{h} = \begin{bmatrix} 0.52 \\ 0.85 \\ 0 \\ 0 \end{bmatrix}$$

We can see that node 3 has the highest authority weight because it has 2 incoming links, where one of them is coming from the best hub node 2. Node 1 is still better authority than 4 in spite of the same number of incoming links, but this is because node 1 is reinforced by node 2, which is the highest hub and node 4 has got link from node 3 which has no hub value. Now we change our web graph a little:

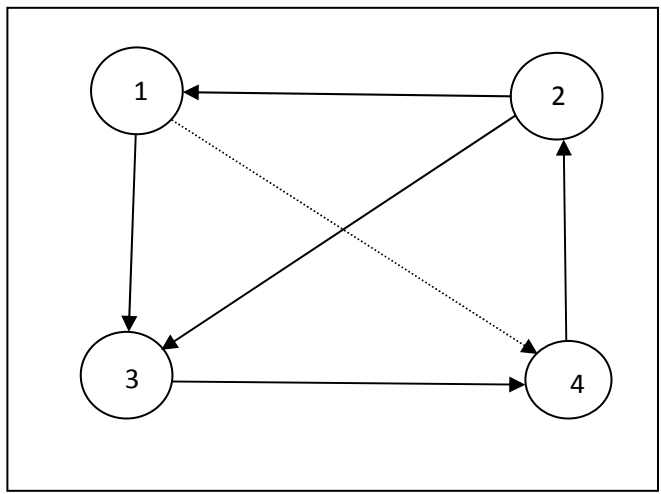

**Figure 5**

We have added a link from node 1 to node 4. The new adjacent matrix, denoted by $\mathbf{B}$, is:

$$\mathbf{B} = \begin{bmatrix} 0 & 0 & 1 & 1 \\ 1 & 0 & 1 & 0 \\ 0 & 0 & 0 & 1 \\ 0 & 1 & 0 & 0 \end{bmatrix}$$

One can see that we have added number 1 in the first row fourth column to identify a link from node 1 to node 4. The results of the Eigenanalysis we obtain after this small alteration are:

For authority weights:

$$\mathbf{a}'' = \begin{bmatrix} 0.32 \\ 0 \\ 0.73 \\ 0.59 \end{bmatrix}$$

And for hub weights are:

$$\mathbf{h}'' = \begin{bmatrix} 0.73 \\ 0.59 \\ 0.32 \\ 0 \end{bmatrix}$$

Now node 1, which has got an equal number of outgoing links to node 2 has become the node with the highest hub value, because it connects to 3 and 4 which have higher authority values than nodes 1 and 4, which connect to node 2. Node 4 has received a higher authority value than node 1 because of a link from the better hub – node 1.  Therefore, we can see that HITS gives higher authority weights to those nodes which connect to better hubs and higher hub weights to the nodes connecting to better authorities. Now, having explained in quite some detail how HITS's work, we can move on to analysis of teams in the English premier League.

## Analysis of the English Premier League season 2010-2011

Before we proceed with the actual analysis we shall explain notations and demonstrate details using a toy example. Let us assume the following mini league with only four teams A, B, C and D where, after the first round of games, the following results were observed[4]:

| Teams | Points |
|-------|--------|
| A     | 6      |
| B     | 3      |
| C     | 3      |
| D     | 6      |

**Table 1. Mini league results**

From Table 1 we can see that two teams registered two wins and one loss and two teams registered one win and two losses. The results distributed as follows:

- Team A beaten teams B and C and lost to team D
- Team B beaten team C and lost to teams D and A
- Team C beaten team D and lost to teams B and A
- Team D beaten teams B and A and lost to team C

Now we have to convert these results into a graph, but first let us agree on notation. The following graph means that team A beat team B, link is weighted by 3 points in adjacency matrix.

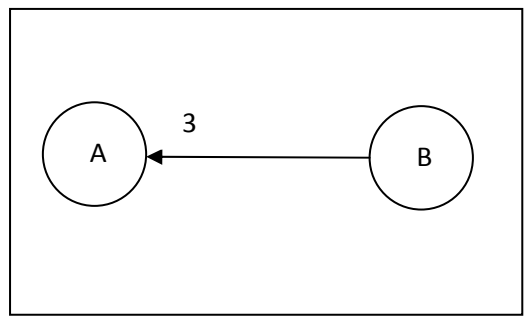

**Figure 6 Team B lost to team A**

In web language it means that node B acknowledges authority of node A and links to it. Now as we remember from the previous part a good hub increases the authority weight by linking to it and a good authority reciprocally increases the hub weight. Hence, this notation suggests that hub ranking will depict losing teams and authority ranking will depict winning teams. If a draw is registered we shall adopt the following pattern:

---

[4] For now we don't bother with goal scores and home and away games.

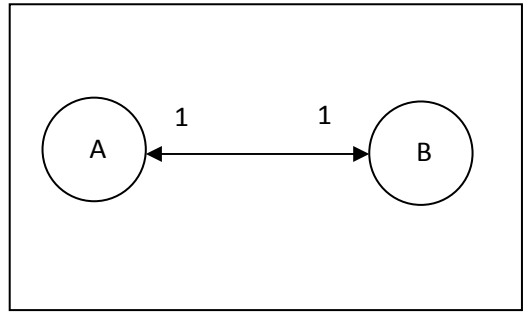

**Figure 7**

In web language it means that both teams acknowledge authority of each other and link reciprocally, and have one point each in the adjacent matrix.

In the first section we explained <in the actual algorithm> that page authority weight grows if pages with higher hub weights link to it, but in the case of football competitions, higher hub weight signifies the losing team, because graph edge points out from losing team to winning team. Hence beating the frequently losing team will influence authority rank more than beating a never losing team, which of course will have a very low, or zero hub weight. It seems reasonable to assume that team desire to increase authority weight will adopt the strategy of beating the weak teams and at least drawing with stronger teams. Therefore, in the case of authority ranking we should expect results somewhat similar to official results. However things are not that straightforward, when hub ranking is considered. Please recall that page will have a high hub weight, if it links to pages with high authority. In our football terms it means that team that lost, or drawn a game to a team with smallest authority weight will have smallest hub weight. A team losing to a team with high authority rank will have higher hub weight, and out of two teams with equal number of wins, losses and draws one which lost more games to high authority rank teams will have higher hub weight. Hence, the lower hub weight is, the better a team is. The strategy team should adopt, is try to beat the team with high authority rank, because losing a game to the team with high authority rank increases ones hub rank significantly. Hence, unlike, in the current system, when simply points calculated hubs and authority weights will take into account, quality[5] of opposition. Based on the authority and hub rankings we can evaluate losses and wins of the teams by looking at the performance as a whole. In mathematical language we would say by describing a problem in multiple dimensions, while the current point system actually reduces the problem to two dimensions.

Now when we defined basic notations for our graph we are ready to draw it (see Figure 8) and calculate adjacent matrix based on the diagram of our mini league results.

---

[5] Quality of a team will of course depend on which teams it beats during the course of the competition.

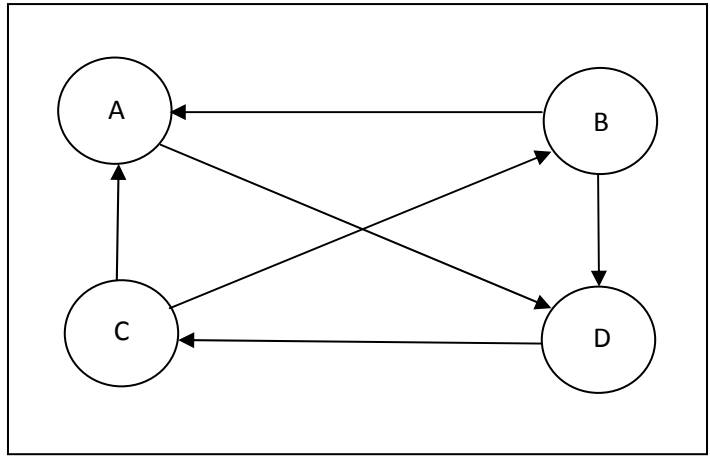

**Figure 8 Graph of mini league results**

The following is the adjacent matrix of the graph:

$$\mathbf{A} = \begin{bmatrix} 0 & 0 & 0 & 3 \\ 3 & 0 & 0 & 3 \\ 3 & 3 & 0 & 0 \\ 0 & 0 & 3 & 0 \end{bmatrix}$$

After performing Eigenanalysis we obtain the following weights for authority:

$$\mathbf{a} = \begin{bmatrix} 0.73 \\ 0.32 \\ 0 \\ 0.59 \end{bmatrix}$$

And for hub:

$$\mathbf{h} = \begin{bmatrix} 0.32 \\ 0.73 \\ 0.59 \\ 0 \end{bmatrix}$$

Let us start from discussing the hub weights, which should bring some interesting insights on the team's performance. Here the best team, the worst hub is actually team D with hub weight 0, which managed to lose (give link) to team C, which in turn is the worst authority. Team A takes the second position in our results with hub weight 0.32. This result can be explained by the fact that team A has lost a game to team D, which is the highest authority after A itself. Team B has lost two games to the strongest teams, D and A , hence takes last position. In web language it would mean that it reinforced WebPages having the highest authority. We can see that hub weights express relationships between teams more comprehensively than regular points system by taking into account not only their results, but looking at their performance as a whole. Hence, using HITS one wouldn't see a team become champion, by losing to the strongest teams and beating only weak teams.

Now we can move to the authority results. In spite of teams A and D sharing an equal number of points, team A has got higher authority weight. It can be explained by the fact that although team A lost to D, it still managed to beat C, which in turn managed to beat D. Team B has got a higher score than team C simply because B managed to beat C.

According to [1] hubs and authorities exhibit what could be called a mutually reinforcing relationship: a good hub is a page that points to many authorities, which is a losing team in our case, and a good authority is a page that is pointed to by many good hubs. Hence, losing to a team with high authority should increase hub weight more significantly than losing to a team with low authority. This supports what we have said above; one can't be champion without beating the most important teams.

Now after all the details of the analysis have been explained we can start to look at the results of the English Premier League (EPL). The results we will be considering are from the 2010-2011 season. We shall not replicate the results of all the games of the season here as they can be easily found in [6], but just show the final standing and points scored by teams at the end of the season (see Table 2).

| Pos | Team | W | D | L | GF | GA | GD | Pts |
|---|---|---|---|---|---|---|---|---|
| 1 | Manchester United | 23 | 11 | 4 | 78 | 37 | 41 | **80** |
| 2 | Chelsea | 21 | 8 | 9 | 69 | 33 | 36 | **71** |
| 3 | Manchester City | 21 | 8 | 9 | 60 | 33 | 27 | **71** |
| 4 | Arsenal | 19 | 11 | 8 | 72 | 43 | 29 | **68** |
| 5 | Tottenham Hotspur | 16 | 14 | 8 | 55 | 46 | 9 | **62** |
| 6 | Liverpool | 17 | 7 | 14 | 59 | 44 | 15 | **58** |
| 7 | Everton | 13 | 15 | 10 | 51 | 45 | 6 | **54** |
| 8 | Fulham | 11 | 16 | 11 | 49 | 43 | 6 | **49** |
| 9 | Aston Vila | 12 | 12 | 14 | 48 | 59 | -11 | **48** |
| 10 | Sunderland | 12 | 11 | 15 | 45 | 56 | -11 | **47** |
| 11 | West Bromwich Albion | 12 | 11 | 15 | 56 | 71 | -15 | **47** |
| 12 | Newcastle United | 11 | 13 | 14 | 56 | 57 | -1 | **46** |
| 13 | Stoke City | 13 | 7 | 18 | 46 | 48 | -2 | **46** |
| 14 | Bolton Wanderers | 12 | 10 | 16 | 52 | 56 | -4 | **46** |
| 15 | Blackburn Rovers | 11 | 10 | 17 | 46 | 59 | -13 | **43** |
| 16 | Wigan Athletics | 9 | 15 | 14 | 40 | 61 | -21 | **42** |
| 17 | Wolverhampton Wanderers | 11 | 7 | 20 | 46 | 66 | -20 | **40** |
| 18 | Birmingham City | 8 | 15 | 15 | 37 | 58 | -21 | **39** |
| 19 | Blackpool | 10 | 9 | 19 | 55 | 78 | -23 | **39** |
| 20 | West Ham United | 7 | 12 | 19 | 43 | 70 | -27 | **33** |

Table 2 Final results of EPL for 2010-2011

The next step will be to create adjacency matrix based on the games results (see Table 3):

| | Arsenal | Aston Vila | Birmingham City | Blackburn Rovers | Blackpool | Bolton Wanderers | Chelsea | Everton | Fulham | Liverpool | Manchester City | Manchester United | Newcastle United | Stoke City | Sunderland | Tottenham Hotspur | West Bromwich Albion | West Ham United | Wigan Athletics | Wolverhampton Wanderers |
|---|---|---|---|---|---|---|---|---|---|---|---|---|---|---|---|---|---|---|---|---|
| Arsenal | 0 | 3 | 0 | 1 | 0 | 0 | 0 | 0 | 0 | 1 | 1 | 0 | 3 | 0 | 1 | 3 | 3 | 0 | 0 | 0 |
| Aston Vila | 3 | 0 | 1 | 0 | 0 | 1 | 1 | 0 | 1 | 0 | 0 | 1 | 0 | 1 | 3 | 3 | 0 | 0 | 1 | 3 |
| Birmingham City | 3 | 1 | 0 | 0 | 0 | 0 | 0 | 3 | 3 | 1 | 1 | 1 | 3 | 0 | 0 | 1 | 3 | 1 | 1 | 1 |
| Blackburn Rovers | 3 | 0 | 1 | 0 | 1 | 0 | 3 | 0 | 1 | 0 | 3 | 1 | 1 | 3 | 1 | 3 | 0 | 1 | 0 | 0 |
| Blackpool | 0 | 1 | 3 | 3 | 0 | 0 | 3 | 1 | 1 | 0 | 3 | 3 | 1 | 1 | 3 | 0 | 0 | 3 | 3 | 3 |
| Bolton Wanderers | 0 | 0 | 1 | 0 | 1 | 0 | 3 | 0 | 1 | 3 | 3 | 1 | 0 | 0 | 3 | 0 | 0 | 0 | 1 | 0 |
| Chelsea | 0 | 1 | 0 | 0 | 0 | 0 | 0 | 1 | 0 | 3 | 0 | 0 | 1 | 0 | 3 | 0 | 0 | 0 | 0 | 0 |
| Everton | 3 | 1 | 1 | 0 | 0 | 1 | 0 | 0 | 0 | 0 | 1 | 3 | 0 | 0 | 0 | 3 | 1 | 1 | 1 | |
| Fulham | 1 | 1 | 1 | 0 | 0 | 0 | 1 | 1 | 0 | 3 | 3 | 1 | 0 | 0 | 1 | 3 | 0 | 3 | 0 | 0 |
| Liverpool | 1 | 0 | 0 | 0 | 3 | 0 | 0 | 1 | 0 | 0 | 0 | 0 | 0 | 1 | 3 | 0 | 0 | 1 | 3 | |
| Manchester City | 3 | 0 | 1 | 1 | 0 | 0 | 0 | 3 | 1 | 0 | 0 | 1 | 0 | 0 | 0 | 0 | 0 | 0 | 0 | 0 |
| Manchester United | 0 | 0 | 0 | 0 | 0 | 0 | 0 | 0 | 0 | 0 | 0 | 0 | 0 | 0 | 0 | 0 | 1 | 0 | 0 | 0 |
| Newcastle United | 1 | 0 | 0 | 3 | 3 | 1 | 1 | 3 | 1 | 0 | 3 | 1 | 0 | 3 | 0 | 1 | 1 | 0 | 1 | 0 |
| Stoke City | 0 | 0 | 0 | 0 | 3 | 0 | 1 | 0 | 3 | 0 | 1 | 3 | 0 | 0 | 0 | 3 | 1 | 1 | 3 | 0 |
| Sunderland | 1 | 0 | 1 | 0 | 3 | 0 | 3 | 1 | 3 | 3 | 0 | 1 | 1 | 0 | 0 | 3 | 3 | 0 | 0 | 3 |
| Tottenham Hotspur | 1 | 0 | 0 | 0 | 1 | 0 | 1 | 1 | 0 | 0 | 1 | 1 | 0 | 0 | 1 | 0 | 1 | 1 | 3 | 0 |
| West Bromwich Albion | 1 | 0 | 0 | 3 | 0 | 1 | 3 | 0 | 0 | 0 | 3 | 3 | 0 | 3 | 0 | 1 | 0 | 1 | 1 | 1 |
| West Ham United | 3 | 3 | 3 | 1 | 1 | 3 | 3 | 1 | 1 | 0 | 3 | 3 | 3 | 0 | 3 | 0 | 1 | 0 | 0 | 0 |
| Wigan Athletics | 1 | 3 | 0 | 0 | 3 | 1 | 3 | 1 | 1 | 1 | 3 | 3 | 3 | 1 | 1 | 1 | 0 | 0 | 0 | 0 |
| Wolverhampton Wanderers | 3 | 3 | 0 | 3 | 0 | 3 | 0 | 3 | 1 | 3 | 0 | 0 | 1 | 0 | 0 | 1 | 0 | 1 | 3 | 0 |

**Table 3 Adjacency matrix based on the results of EPL 2010-2011 season**

| Team | Authority Weights |
|---|---|
| Manchester City | 0.342 |
| Chelsea | 0.328 |
| Manchester United | 0.303 |
| Arsenal | 0.296 |
| Tottenham Hotspur | 0.267 |
| Newcastle United | 0.231 |
| Sunderland | 0.226 |
| Blackpool | 0.211 |
| Fulham | 0.210 |
| Everton | 0.204 |
| Aston Villa | 0.200 |
| Wigan Athletic | 0.197 |
| Blackburn Rovers | 0.179 |
| Liverpool | 0.176 |
| Birmingham City | 0.166 |
| Wolverhampton Wanderers | 0.164 |
| West Bromwich Albion | 0.163 |
| West Ham United | 0.148 |
| Stoke City | 0.146 |
| Bolton Wanderers | 0.139 |

**Table 4 Authority ranking**

| Team | Hub Weights |
|---|---|
| Manchester United | 0.008 |
| Chelsea | 0.087 |
| Manchester City | 0.111 |
| Tottenham Hotspur | 0.133 |
| Liverpool | 0.135 |
| Everton | 0.160 |
| Arsenal | 0.166 |
| Bolton Wanderers | 0.203 |
| Aston Villa | 0.205 |
| Fulham | 0.214 |
| Stoke City | 0.215 |
| Wolverhampton Wanderers | 0.238 |
| West Bromwich Albion | 0.241 |
| Birmingham City | 0.241 |
| Newcastle United | 0.243 |
| Blackburn Rovers | 0.266 |
| Sunderland | 0.272 |
| Wigan Athletic | 0.307 |
| Blackpool | 0.338 |
| West Ham United | 0.362 |

Table 5 Hub ranking

As we can see the order of the teams in our tables (see Table 4 for authority weights and Table 5 for hub weights) is different to the official results for the 2010-2011season in the EPL. For instance the *official table* has got the top four teams in the following order: Manchester United, Chelsea, Manchester City and Arsenal. The *authority table* has got the top four teams in this order: Manchester City, Chelsea, Manchester United and Arsenal. In the *hub weights table the* top four[6] are: Manchester United, Chelsea, Manchester City and Tottenham Hotspur. Thus, for instance, we can see that Arsenal although it has an equal number of losses to Tottenham but has a higher hub index than them. This can be explained by the fact that Tottenham lost their games to teams with smaller authority rankings, which makes it a worse hub, but a better team. In web language, the Tottenham page has outgoing links to web pages with smaller authority weights than the Arsenal page.

## Conclusion

We have demonstrated the application of HITS algorithm to the problem of ranking teams in the English Premier League. The algorithm allows the ranking of teams using the notions of hubs and authorities well known for ranking pages in the World Wide Web. In application to our problem we have defined authority as a team which wins games and hub as a team which loses games. Using this convention we can rank teams not only according to the number of wins and losses, but according to the quality of opposition and outcome of the game. Using this ranking system, the champion's team, which is a 'worst hub' must not lose games to other 'good authorities' teams. If by the end of the competition there are teams with an equal number of wins and losses then the team which has beaten more teams with higher authority ranks, wins. Hence, a team can't become a champion by beating only weak teams and drawing and losing to strong teams. It's important to note that in this work we didn't look at goal difference and only took into account losses and wins. However as a future research direction it might be interesting to devise a weighting system which takes into account not only losses and wins, but goal difference and for instance home and away games too. Such a ranking system would be the most comprehensive way to evaluate the performance of football teams.

---

[6] Please remember that in our analysis small value for hub weight is actually a good thing

Reference

1. Kleinberg, J., *Authoritative sources in a hyperlinked environment.* Journal of the Association for Computing Machinery, 1999. **46**(5): p. 604-632.
2. Page, L., et al., *The PageRank Citation Ranking: Bringing Order to the Web*. 1999, Stanford InfoLab.
3. Dumais, S.T., *Latent Semantic Analysis.* Annual Review of Information Science and Technology, 2004. **38**(1): p. 188-230.
4. Strang, G., *Linear Algebra and Its Applications*. 1988: Thomson Brooks/Cole.
5. Wilkinson, J.H., *The algebraic eigenvalue problem*. 1988: Clarendon Press.
6. Wikipedia. *2010–11 Premier League*. 2011 [cited; Available from: http://en.wikipedia.org/wiki/2010%E2%80%9311_Premier_League.